\documentclass{PoS}

\title{Swift, UVOT and Hot Stars}

\ShortTitle{UVOT Hot Stars}

\author{\speaker{Michael H. Siegel}\\
        The Pennsylvania State University\\
	Department of Astronomy and Astrophysics\\
        E-mail: \email{siegel@swift.psu.edu}}

\author{Caryl A. Gronwall\\
        The Pennsylvania State University\\
	Department of Astronomy and Astrophysics\\
        E-mail: \email{caryl@astro.psu.edu}}

\author{Lea M. Z. Hagen\\
          The Pennsylvania State University\\
 	Department of Astronomy and Astrophysics\\
         E-mail: \email{lmz5057@psu.edu}}

\author{Erik A. Hoversten\\
        University of North Carolina at Chapel Hill\\
 	Department of Physics and Astronomy\\
       E-mail: \email{ehoverst@live.unc.edu}}

\abstract{We present the results of our ongoing investigation into the properties of hot stars and young
stellar populations using the Swift/UVOT telescope. We present UVOT photometry of open and globular
clusters and show that UVOT is capable of characterizing a
variety of rare hot stars, including Post-Asymptotic Giant Branch and Extreme Horizontal Branch Stars.
We also present very early reults of our survey of stellar populations in the Small Magellanic Cloud.
We find that the SMC has experienced recent bouts of star formation but constraining the exact star
formation history will depend on finding an effective model of the reddening within the SMC.
          }

\FullConference{Swift: 10 Years of Discovery,\\
		2-5 December 2014\\
		La Sapienza University, Rome, Italy }

\begin{document}

\section{Background}

Hot stars, defined as those with effective temperatures greater than 7000 degrees, are the final frontier in stellar evolution.  While
most phases of stellar evolution are well-understood due to copious observational data and well-constrained theoretical modelling,
the properties and underlying astrophysics of hot stars remain contentious (see a discussion, in the context
of Swift, in Siegel et al. 2014).  Of particular importance are
the astrophysics of massive stars, which play fundamental roles in chemical enrichment, galaxy evolution, core-collapse supernovae and 
GRBs (see, e.g., Langer et al. 2012).
Hot stars and young stellar populations are best studied in the ultraviolet (UV), where their spectral energy distributions
peak and the contributions from cool stars and older stellar populations is minimized.

\begin{figure}[h!]
\includegraphics[width=14cm]{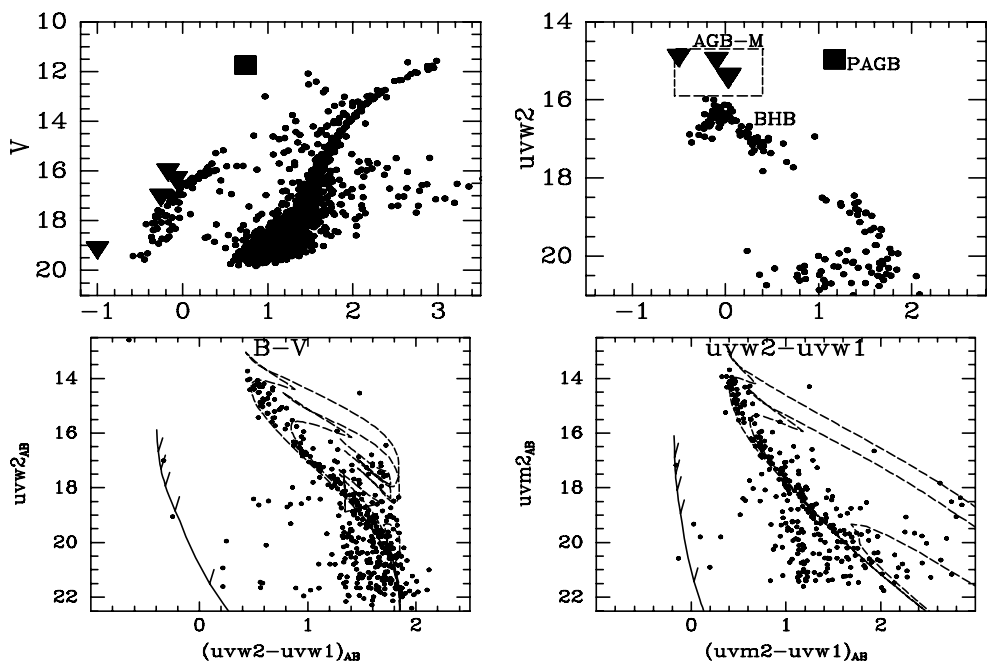}
\caption{Hot stars in clusters.  The top panels show the optical and NUV color-magnitude diagrams for the
old stellar populations in the M~79 globular cluster.  The UV-bright stellar
populations are labeled.  M~79's red giant branch (RGB) dominates the optical but is too cool and thus to faint in the
UV to be detected by Swift/UVOT.  By contrast, the BHB, EHB, PAGB (square) and AGB-M (triangles) stand
out. The bottom panels show the color-magnitude diagrams of a young stellar population
in the NGC~2539 open cluster.  Isochrones  from the Padova group (Bressan et al. 2012)
are overlaid with parameters set to $\rm[Fe/H]=+0.0$, $E(B-V) = 0.05$,
$(m-M)_0=10.65$, and ages of 630 Myr, 1 Gyr and 5 Gyr. The solid blue line is a
$0.5\, M_{\circ}$ white dwarf sequence using cooling curves from Bergeron et al.\ (2011).}
\label{f:twoclus}
\end{figure}

Figure \ref{f:twoclus} shows the variety of stars that are detected in  the UV.  The color-magnitude
diagrams (CMD) show two stellar populations -- one old and one young.  The old globular cluster M~79 shows a clear blue horizontal branch (BHB).
The bend in the BHB corresponds to the extreme horizontal branch (EHB), helium-burning stars with temperatures over 16,000 K that have
extremely thin envelopes and will likely evolve directly to the white dwarf sequence.  These stars play a role in the ``UV upturn" seen
in old elliptical galaxies (see, e.g.,  O'Connell 1999, Sandquist et al. 2008).  The CMD also shows a Post-Asymptotic Giant Branch star (PAGB),
a star in the rare short phase of evolution from the tip of the asymptotic giant branch to the tip of the white dwarf sequence (Bond 1997).
These stars may be standard candles.  Also revealed in the NUV are rare AGB Manque stars (AGB-M). The nature of these
stars is unclear but they may be the descendants of EHB stars that have envelopes too small to read the AGB (Greggio \& Renzini 1990).
NGC~2539, a young open cluster, shows a young blue main
sequence and a sparse sequence of hot white dwarfs.

Studying hot stars in the UV has implications beyond stellar astrophysics.  In studies of extragalactic objects, 
the light of unresolved stellar populations can reveal the properties of the stars and the dust that comprise the unresolved
populations, unveiling their complex formation history (see, e.g., Hoversten et al. 2011).  However, the spectral synthesis models
used in the ultraviolet are limited
by a lack of empirical data on young stellar populations (Bruzual 2009).  By measuring the integrated light of nearby young
stellar populations -- populations which can be studied on a star-by-star basis -- we can provide the emperical baseline needed to constrain
the models of stellar populations.

The Swift/UVOT instrument is uniquely suited to studying hot stars.  It has a wide (17') field of view, which enables the study
of entire clusters and large numbers of field stars, providing access to complete stellar populations.  Its NUV sensitivity
gets further into the spectral energy distribution of hot stars, providing better leverage on their bulk properties.  Finally,
the 2.3 arc-second resolution allows us to resolve stars in cluster cores and nearby galaxies.  These advantages are particularly
notable in the case of Galactic open clusters, which are sometimes too large and sparse to be effectively studied with HST
and too close to the Galactic midplane to be have been observed by GALEX.

In this contribution, we highlight the results of an ongoing survey of nearby hot stars designed to address multiple astrophysical
issues.  Our survey has studied field white dwarf stars (Siegel et al. 2011, 2013), open and globular clusters (Siegel et al. 2014), the
nearby Magellanic Clouds and will soon expand to M31.

\section{Open and Globular Clusters}

Over the last several years, we have been performing an imaging survey of Galactic star cluters.  Open clusters generally provide
information on young stellar populations while globular clusters provide information on older stellar populations  Studying these
objects not only provides information on the rare phases of stellar evolution described above, it provides the empirical baseline
needed to study unresolved stellar populations in more distant galaxies.

To date, we have studied 97 open clusters and 65 globular clusters, typically obtaining at least 1 ks in each of UVOT's
three NUV filters (uvw1, uvm2 and uvw2).  The observational aspects are discussed in detail in Siegel et al. (2014).

\begin{figure}
\includegraphics[width=14cm]{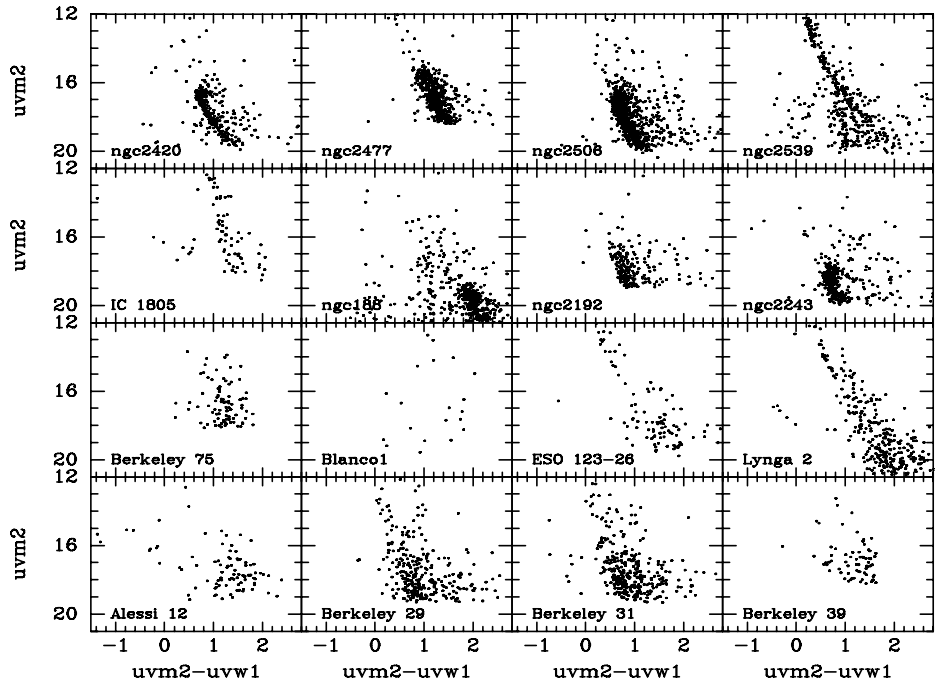}
\caption{A sampling of open cluster color-magnitude diagrams produced from Swift/UVOT data.  The clusters span a broad range
of age, metallicity and reddening.}
\label{f:allopen}
\end{figure}

\begin{figure}
\includegraphics[width=14cm]{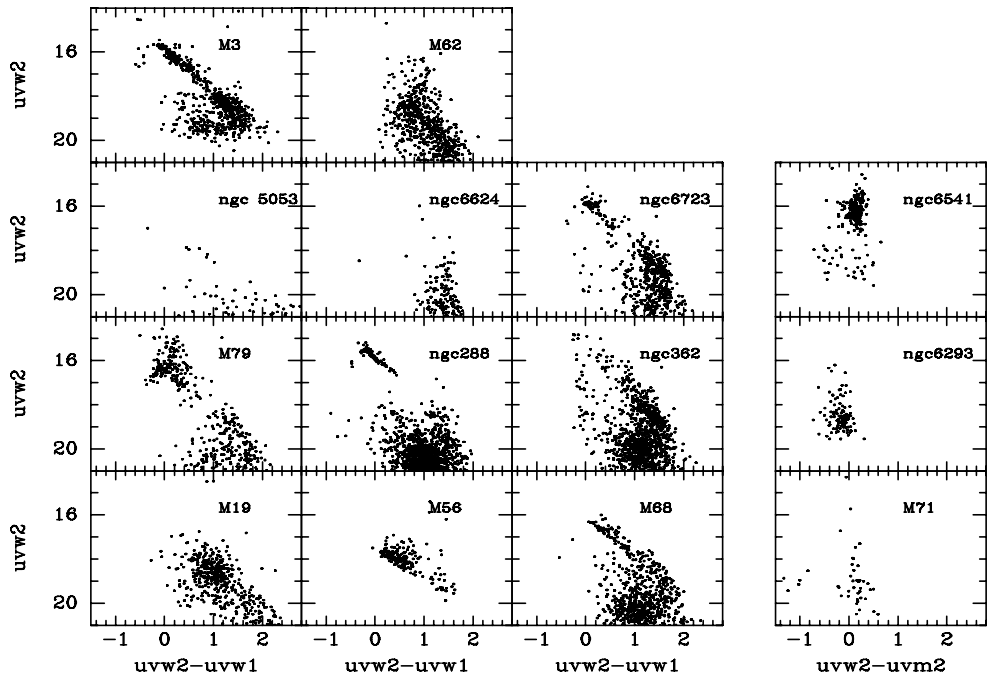}
\caption{A sampling of globular cluster color-magnitude diagrams produced from Swift/UVOT data.  The clusters span a broad range
of age, metallicity and reddening.}
\label{f:allglob}
\end{figure}

Figures \ref{f:allopen} and \ref{f:allglob} show a sampling of the CMDs. Most show the the distinctive features seen in Figure \ref{f:twoclus}.
For the globular clusters, this provides a unique dataset of PAGB, AGB-M and EHB stars.  At present, we have not
detected any Blue Hook stars, a rare extremely hot (35,000 K) class of star that have only been postively identified in the most massive 
globular clusters from HST (see, e.g., Brown et al. 2012).

Many of the open clusters show clear main sequences, providing empirical constraint on the properites of intermediate and high-mass ($>8 M_{\circ}$)
stars.  However, a significant fraction of the open clusters do not show a clear main sequence due to their small size or differential reddening
in the field.  These clusters are, however, very useful for the study of the integrated light of stellar populations.

\section{SUMAC}

The Swift UVOT MAgellanic Clouds Survey (SUMAC) was an extensive program to observe the central regions of
both Magellanic Clouds with the Swift XRT and UVOT.  The imaging took place between 2010 September 26 and 2013 November 5.
For the LMC, it comprised 2200 images in 150 fields with a typical exposure time of 3 ks in UVOT's three NUV filters
for a total exposure time of 5.4 days.  For the SMC, the survey obtained 656 images in 50 fields with a tpyical exposure time of 3 ks
in UVOT's three NUV filters for a total exposure time of 1.8 days.
The images were processed, stacked, photometered and calibrated according to the method outlined in Siegel et al. (2014).  We detected
approximately 250,000 UV souces in the SMC and nearly a million in the LMC.
Full-color
mosaic images are available at the NASA website.  Processed
files and point source catalog will be made publicly available later this year.

\begin{figure}
\includegraphics[width=14cm]{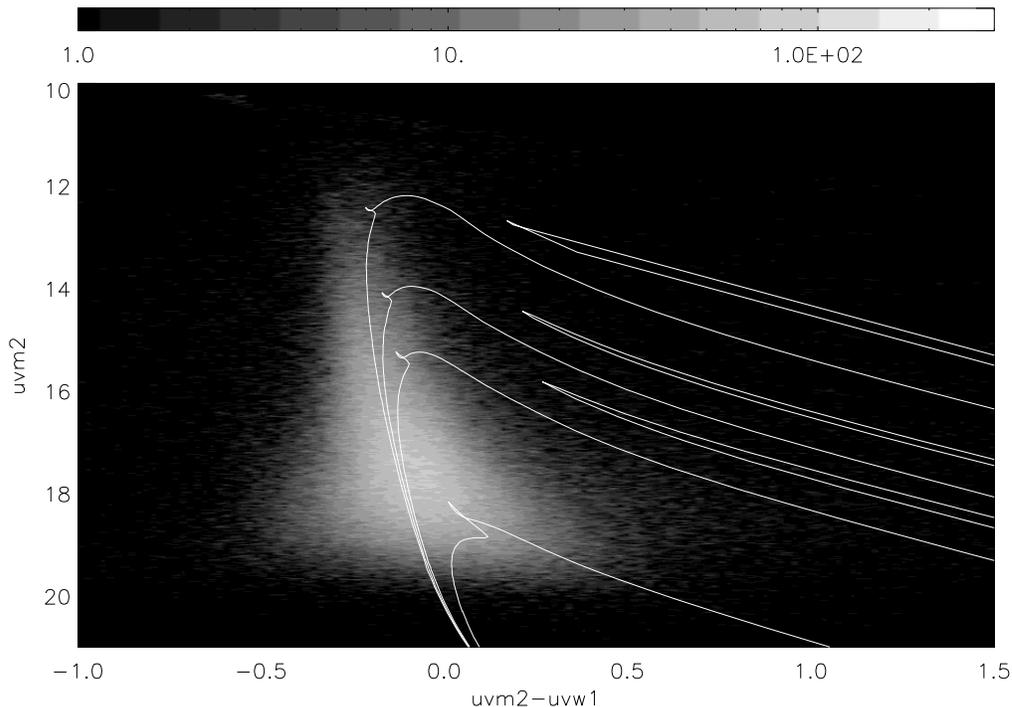}
\caption{A Hess diagram of the uvm2-uvw1 photometry for 250,000 point sources measured in the SUMAC survey.  Overlayed for comparison
are isochrones from the Padova program set to [Fe/H]=-0.5, E(B-V)=.05 and ages of 20, 50, 100 and 500 Myr..}
\label{f:smcbig}
\end{figure}

Figure \ref{f:smcbig} shows a Hess diagram of the complete photometric sample from the SMC with comparison stellar populations overlaid.
Optical CMDs of the SMC show two distinct sequences -- the blue plume of hot main sequence
stars from the younger stellar populations and a very strong red giant branch (RGB) of older stellar populations
(see Harris \& Zaritsky 2004, herafter HZ04).  However, the RGB stars are too cool to be detected with UVOT.  This allows SUMAC
to probe the last 500 Myr of star formation with unprecedented detail and clarity.
As can be seen, the UVOT data is sensitive to the last 500 Myr of star formation.  

To probe the recent star formation history of the SMC, we utilized the StarFISH star formation history program of HZ04 to deconvolve
the CMDs into their component stellar populations.  Isochrones and luminosity functions were taken from the Padova program.
For the purpose of this first effort, we assumed that the reddening law of the
SMC was the "SMC law" from Pei et al. (1992), which does not have a blue bump at 2175 \AA.
After numerous tests, we found that
the reddening distribution that best fit the data was the "hot star" distribution from HZ04.

\begin{figure}
\includegraphics[width=14cm]{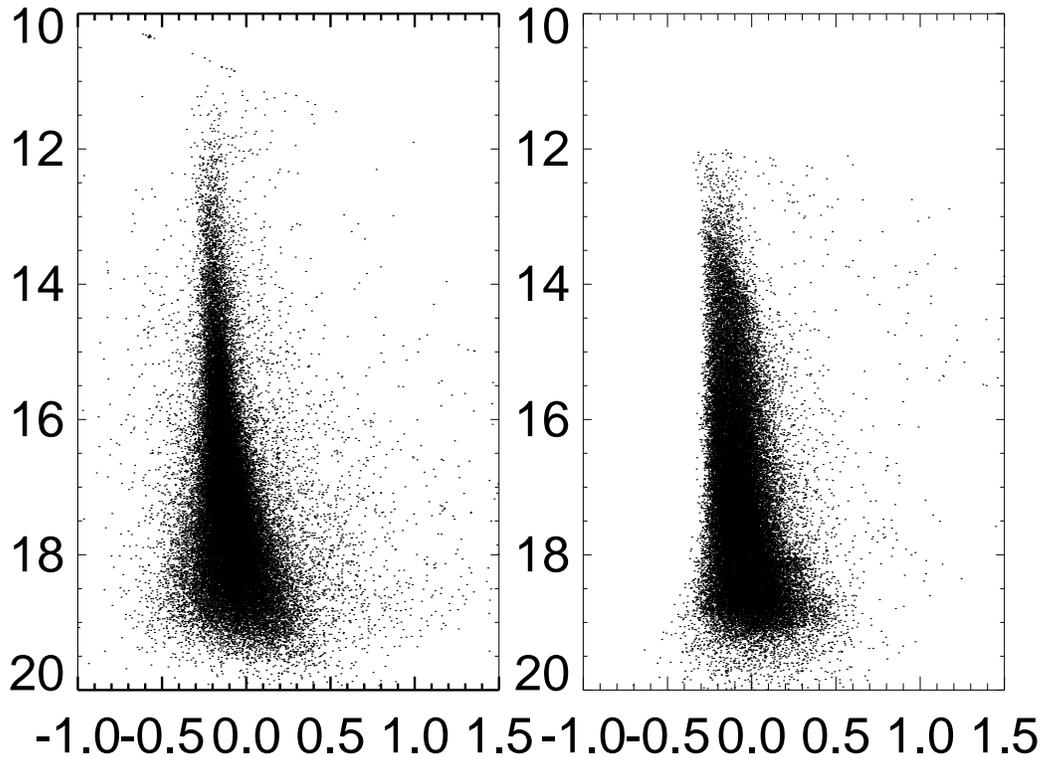}
\caption{A comparison of the UVOT observational data (left) to the synthetic photometry produced by StarFISH (right) for a sample of ten
UVOT fields.  As can be seen, the
salient feature of the CMD -- the long blue plume -- is well-reproduced by the simulation.  The synthetic data does show a larger
spread in color space likely due to cruder binning of the CMDs and an overly pessimistic error model.}
\label{f:simstars}
\end{figure}

\begin{figure}
\includegraphics[width=14cm]{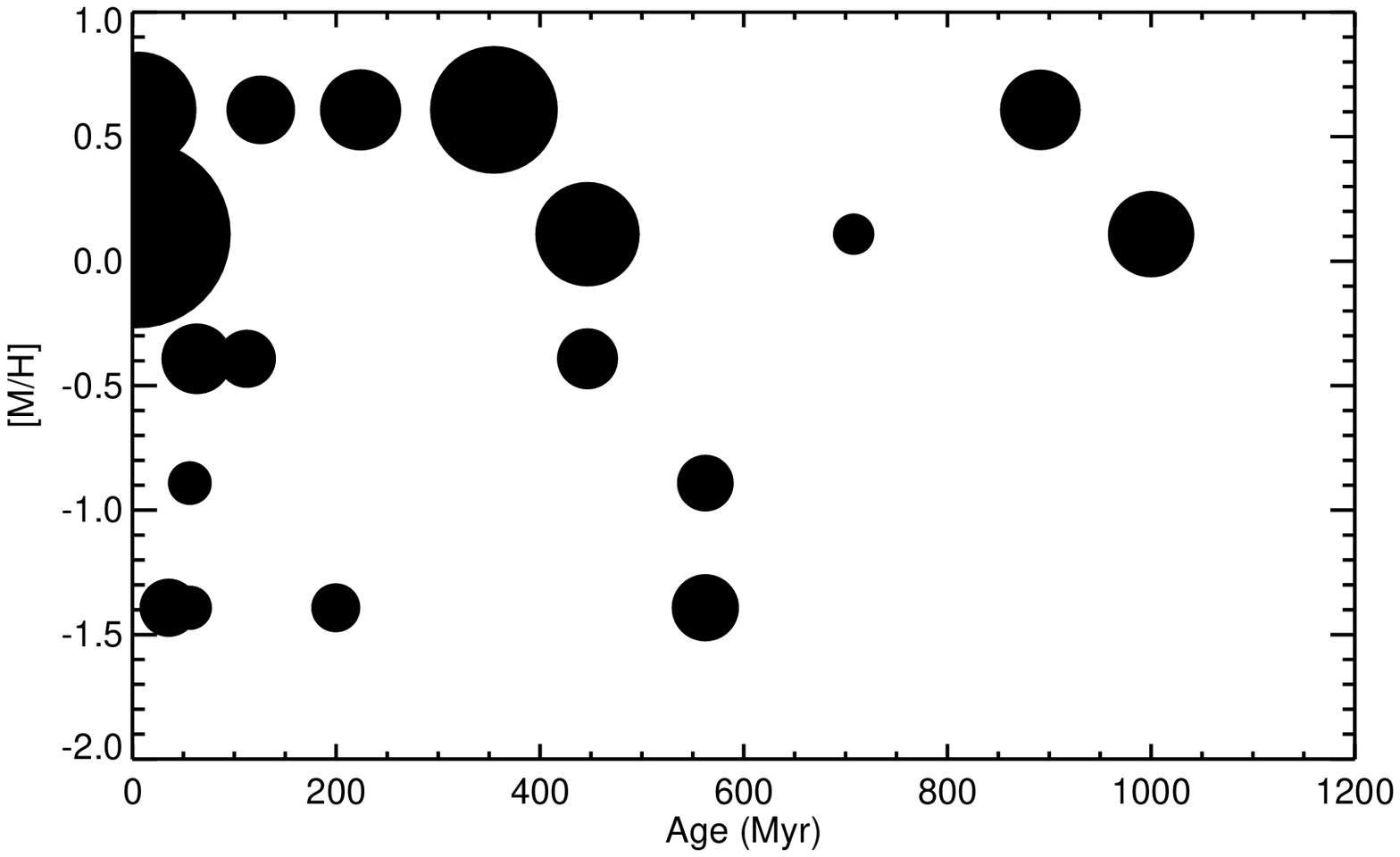}
\caption{The star-formation history used to produce the color-magnitude diagram above.  Points are stellar populations that are included
in the CMD, with the size of the point indicating the overall star-formation rate.  The simulations shows a number of
interesting features, including a young burst of star formation at about 400 Myr and a very young burst of star formation at a few Myr.  The 
stellar population are more metal-rich and younger than expected, both of which are likely a result of a poor reddening model.}
\label{f:sfh}
\end{figure}

Figure \ref{f:simstars} shows a comparison of the UVOT photometry from ten of our SUMAC field to the first run of simulated StarFISH photometry.
As can be seen, the salient features of the CMD -- 
specifically the long narrow blue plume of young stars -- is easily reproduced by the synthetic photometry.  Figure \ref{f:sfh} shows the inferred
star-formation history of these fields.  Our initial results are intriguing but show that more work needs to be done.  The populations are younger
and more metal-rich than expected from previous investigations and spectroscopic studies.  While this may be accurate, it could also reflect
inadequacies in the reddening model used, to which
the UV photometry is particularly sensitive.  We are currently
running simulations with a variety of reddening models in an effort to improve the fit and provide a better comparison of the SMC star formation
history to previous efforts.

Despite this limitation, this preliminary result shows that the UVOT can provided unprecedented detail on the SMC's recent star-formation history.
Furthermore, the final simulations will be run on the individual fields to see how the age, metallicity and reddening law vary over the face of the SMC.
The result will be the most thorough investigation to date of the SMC's recent star formation and a critical test of whether its star formation
history is indeed driven by dynamical interactions with the Milky Way and the Galaxy, as posited by HZ04.

\section{Conclusion}

Our surveys of the star clusters of the Milky Way and the Magellanic Clouds has revealed rich detail in their color-magnitude diagrams and is 
providing an unprecented opportunity to explore the properties of rare hot stars and young stellar populations.  In particular, we are 
obtaining large samples of BHB, EHB, PAGB and AGB-M stars, studying the properties of young main sequence stars and probing
the last 500 Myr of star formation in the SMC.

The final result of this program will not only be a better understanding of the properties of hot stars, but the creation of an empirical
baseline with which to study more distant unresolved stellar populations.  This understanding is particularly critical for young stellar
population in the more distant universe, whose emission peaks in the UV.

Studying the most massive stars (O- and B-type) in the crowded cores
of nearby galaxies is beyond UVOT's capabilities and will require improved future
UV space instrumentation (in particular, sub-arcsecond resolution and multi-filter FUV sensitivity).  Nevertheless, Swift/UVOT is paving
the way for future studies by providing high-quality high-resolution multi-filter NUV
data for thousands of hot stars.

\end{document}